\documentclass[showpacs,preprint]{revtex4}
\usepackage{graphicx,color,epsfig
}

\newcommand{\ba}{\begin{array}}
\newcommand{\ea}{\end{array}}
\newcommand{\bd}{\begin{displaymath}}
\newcommand{\ed}{\end{displaymath}}
\newcommand{\be}{\begin{equation}}
\newcommand{\ee}{\end{equation}}
\newcommand{\beq}{\begin{eqnarray}}
\newcommand{\eeq}{\end{eqnarray}}

\newcommand{\non}{\nonumber\\ }

\begin{document}

\bigskip
\title{Probing new physics in  $B\to J/\Psi~\pi^0$ decay}

\author{Jing-Wu Li$^{1}${\footnote {Email:lijw@xznu.edu.cn}},
~~~Dong-Sheng Du$^{2}${\footnote
{Email:duds@mail.ihep.ac.cn}},~~~Xiang-Yao Wu$^{3}${\footnote
{Email:wuxy2066@163.com}}}

\vspace*{1.0cm}

\affiliation{$^{1}$Department of Physics, Xu Zhou Normal University,
XuZhou 221116, China,\\$^{2}$Institute of High Energy Physics, P.O.
Box 918(4), Beijing 100049,\\$^{3}$Institute of Physics, Jilin
Normal University, Siping 136000, China }


\vspace*{1.0cm}

\date{\today}
\begin{abstract}
We calculate the branching ratio of $B\to J/\Psi~\pi^{0}$  with a
mixed formalism that combines the QCD-improved factorization and the
perturbative QCD approaches. The result is consistent with
experimental data.  The quite small penguin contribution in $B\to
J/\Psi~\pi^{0}$ decay can be calculated  with this method. We
suggest two methods  to extract the weak phase $\beta$. One is
through the dependence of the   mixing induced CP asymmetry
$S_{J/\Psi\pi^{0}}$ on the weak phase$\beta$ , the other is from the
relation of  the total asymmetry $A_{CP}$ with the weak phase
$\beta$. Our result shows that the deviation $ \bigtriangleup
S_{J/\psi\pi^0 }$ of the mixing induced CP asymmetry from
$Sin(-2\beta)$ is of  $ \mathcal{O}(10^{-3})$ and has  much less
uncertainty.  The above $ \mathcal{O}(10^{-3})$ deviation can
provide a good reference for identifying new physics.

\end{abstract}

 \maketitle


B physics is entering the era of  precision measurement, It is not
far from revealing new physics beyond the Standard Model(SM).  Many
authors have studied the topics and suggest some windows for looking
for new physics(NP)\cite{npa1}-\cite{npa9}. Because falvour-changing
neutral current (FCNC) processes only occur at the loop-level in the
SM , so they  are particularly sensitive to NP interactions. It was
pointed out that $B^0_{q}-\bar{B}^0_{q}$ mixing and decays are good
places for new physics to enter through the exchange of new
particles in the box diagrams, or through new contributions at the
tree level \cite{bmixing1}-\cite{bmixing3}, so
$B^0_{q}-\bar{B}^0_{q}$ system has been studied in many papers for
probing new physics\cite{bmixingatt1,bmixingatt5}. $B\to
J/\Psi\pi^0$ decay is a good mode for looking for new physics and
extracting the weak phase $\beta$ . The direct CP asymmetry
$C_{J/\psi\pi^0 }$ and the deviation $\bigtriangleup S_{J/\psi\pi^0
}\equiv S_{J/\psi\pi^0 }-\sin (-2\beta)$ of the mixing -induced CP
asymmetry from $\sin (-2\beta)$ in this decay arise from  quite
small penguin contribution in the SM,  so these quantities  are
sensitive to new physics effect. Comparing  the  prediction of CP
asymmetry in the SM  with the experimental data, one can find new
physics signal.
 Thus it is essential to calculate the $
\bigtriangleup S_{J/\psi\pi^0 }$ and $C_{J/\psi\pi^0 }$ in $B\to
J/\Psi\pi^0$   in the SM accurately.

The deviation $\bigtriangleup S_{J/\psi\pi^0 }=S_{J/\psi\pi^0 }-\sin
(-2\beta)$  or direct CP asymmetry $C_{J/\psi\pi^0 }$ in $B\to
J/\Psi~\pi^0$ decay have been studied in Ref.~\cite{deltasciu} by
fitting to the current experimental data, the result is $
C_{J/\psi\pi^0 }= 0.09\pm0.19$ which has very large uncertainty. In
that case we can not say anything about new physics effects.

     In order to reveal new physics effects, we need both better
 theoretical prediction and experimental measurement with less
 uncertainties. That is the aim of our present paper.

     In what follows, we first evaluate the penguin pollution effect by a method which
have been used to explain  many
 B decays  into charmonia successfully\cite{btocharmn1,btocharmn2}.  We
 find the  penguin pollution in the $B\to J/\Psi~\pi^0$ decay is
 quite   small, the deviation $\bigtriangleup S_{J/\psi\pi^0 }=S_{J/\psi\pi^0 }-\sin (-2\beta)$
 in $B\to J/\Psi~\pi^0$ decay is $\mathcal{O}(10^{-3})$,
  which means that the measured deviation  $\bigtriangleup S_{J/\psi\pi^0 }$ at $1\%$
 will indicate the presence of new physics.

      The latest experimental data of $ \bigtriangleup
S_{J/\psi\pi^0 }$ is $ S_{J/\psi\pi^0}=-0.4\pm0.4$\cite{pdg2006},
which has large  error, so we are expecting to have more
  precise measurement in the near future.

The decay rate of of $B\to J/\Psi~\pi^{0}$ can be written as
\begin{equation}
\Gamma=\frac{1}{32\pi m_B}G_F^2(1-r_2^2+\frac{1}{2}r_{2}^4-r_3^2)
|{\cal A}|^2\;.
\end{equation}
with $r_2=m_{J/\psi}/m_B$, $r_3=m_{\pi}/m_B$.

\begin{figure}[tb]
\begin{center}
\epsfig{file=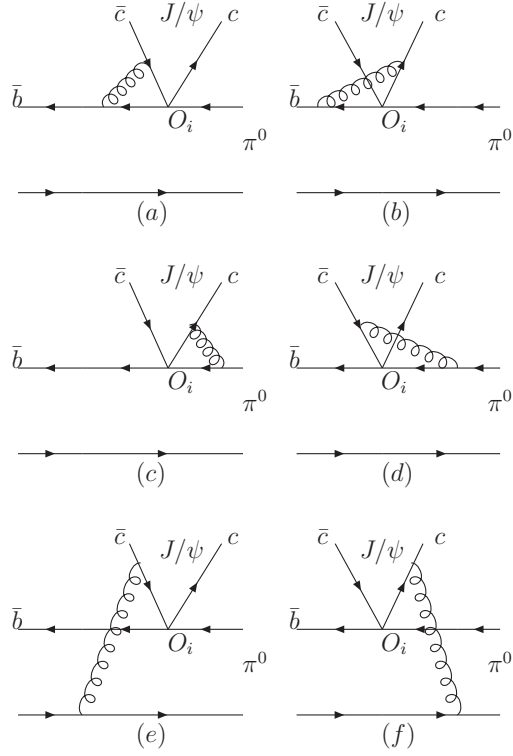,height=10cm}
\caption{ Nonfactorizable contribution to the $B^0\to
J/\psi~\pi^{0}$ decay} \label{nonfc}
\end{center}
\end{figure}

  The amplitude ${\cal A}$ consists of  factorizable part and
  nonfactorizable part.  It can be written as
\begin{eqnarray}
{\cal A}&=&{\cal A}_{NF}+{\cal A}_{VERT}+{\cal A}_{HS}\;,
\label{amppi}
\end{eqnarray}

where ${\cal A}_{NF}$ denote the factorizable contribution in Naive
Factorization Assumption(NF), ${\cal A}_{VERT}$ is the vertex
corrections from Fig.~\ref{nonfc}.(a)-(d) , ${\cal A}_{HS}$ is the
spectator correction from Fig.~\ref{nonfc}.(e)-(f).

The factorizable part ${\cal A}_{NF}$ in Eq.~(\ref{amppi}) for
$B\to J/\Psi \pi^{0}$ decay can not be
 calculated reliably in the pQCD approach,
   because its characteristic scale is around 1 GeV. We
  parameterize the sum of the factorizable part ${\cal A}_{NF}$ and the vertex corrections ${\cal A}_{VERT}$ as,
\begin{eqnarray}
{\cal A}_{NF}+{\cal A}_{VERT}=a_{eff} m_B^2 f_{J/{\psi}}F_1^{B\to
\eta}(m_{J/\psi}^2) (1-r_2^2)\;,\label{amppinf}
\end{eqnarray}
where $f_{J/{\psi}}$ is decay constant of $J/\psi$ meson,

For the $B\to \pi$ transition form factors, we employ the models
derived from the light-cone sum rules \cite{formbtopi}, which have
been parameterized as
\begin{eqnarray}
F_1^{B\to
\pi}(q^2)=\frac{r_1}{1-q^2/m_1^2}+\frac{r_2}{1-q^2/m_{fit}^2}\;\label{f1pi}
\end{eqnarray}

with $r_1=0.744$, $r_2=-0.486$, $m_1=5.32Gev$, $m_{fit}^2=40.73Gev$
for $B \to \pi$ transition.

The factorization and vertex correction from
Fig.~\ref{nonfc}.(a)-(d) can be calculated in the QCDF\cite{qcdf}.
Summing up the factorizable part and vertex correction , we can get
the Wilson coefficient $a_{eff}$,

\begin{eqnarray}
a_{eff}&=&V_c^\ast \left[C_1+V_c^\ast
\frac{C_2}{N_c}+\frac{\alpha_s}{4\pi}\frac{C_F}{N_c}C_2
\left(-18+12\ln\frac{m_b}{\mu}+f_I\right)\right]\nonumber\\&&-V_t^\ast
\Big[C_3+\frac{C_4}{N_c}+\frac{\alpha_s}{4\pi}\frac{C_F}{N_c}C_4
\left(-18+12\ln\frac{m_b}{\mu}+f_I\right)\nonumber\\&&+C_5+\frac{C_6}{N_c}+\frac{\alpha_s}{4\pi}\frac{C_F}{N_c}C_6
\left(6-12\ln\frac{m_b}{\mu}-f_I\right)+C_7+\frac{C_8}{N_c}+C_9+\frac{C_10}{N_c}\Big]\;\label{anfandver}
\end{eqnarray}

with the function,
\begin{eqnarray}
f_I=\frac{2\sqrt{2N_c}}{f_{J/\psi}}\int
dx_3\Psi^L(x_2)\left[\frac{3(1-2x_2)}{1-x_2}\ln x_2-3\pi
i+3\ln(1-r_2^2)+\frac{2r_2^2(1-x_2)}{1-r_2^2 x_2}\right]\;,
\end{eqnarray}

The spectator corrections ${\cal A}_{HS}$ from
Fig.~\ref{nonfc}.(e)-(f), can be calculated reliably in the pQCD as
in Ref.~\cite{btocharmn1,btocharmn2},
\begin{eqnarray}
{\cal A}_{HS}&=&V_c^\ast {\cal M}_{1}^{(J/\psi \pi)}- V_t^\ast {\cal
M}_{4}^{(J/\psi \pi)}-V_t^\ast{\cal M}_{6}^{(J/\psi
\pi)}\;,\label{amppihs}
\end{eqnarray}
where the amplitudes ${\cal M}_{1,4}^{(J/\psi \pi)}$ and ${\cal
M}_{6}^{(J/\psi \pi)}$ result from the $(V-A)(V-A)$ and $(V-A)(V+A)$
operators in the effective Hamiltonian, respectively. Their
factorization formulas are given by the  pQCD approach. In the
calculation of ${\cal M}_{1,4}^{(J/\psi \eta)}$ and ${\cal
M}_{6}^{(J/\psi \eta)}$, because $J/\psi$ is heavy,  we reserve  the
power terms of $r_2$ up  to $\mathcal{O}(r^{4}_{2})$, the power
terms of $r_3$ up to $\mathcal{O}(r^{2}_{3})$ .
\begin{eqnarray}
{\cal M}_{1,4}^{(J/\psi \pi)} &=&16\pi m_B^2 C_{F}\sqrt{2N_{c}}%
\int_{0}^{1}[dx]\int_{0}^{\infty }b_{1}db_{1} \Phi _{B}(x_{1},b_{1})
\nonumber \\
&& \times \Big\{ \Big[ (1-2r^{2}_{2}+r^{4}_{2})(1-x_{2}) \Phi _{\pi}
( x_{3} )\Psi^{L}(x_{2})
+\frac{1}{2}(r^{2}_{2}-r^{4}_{2})  \Phi_{\pi}(x_{3})\Psi^{t }(x_{2}) \nonumber \\
&&- r_{\pi} (1-r^2_2)x_3 \Phi^{p}_{\pi}(x_3) \Psi _{L}(x_{2})+
r_{\pi} \left( 2r^{2}_{2}(1-x_2)+(1-r^2_2)x_3 \right)
\Phi^{t}_{\pi}(x_3) \Psi^{L}(x_{2})
  \Big]\nonumber \\
&&\times E_{1,4}(t_d^{(1)})h_d^{(1)}(x_1,x_2,x_3,b_1)
\nonumber \\%
&& - \Big[ (x_2-x_2r^{4}_{2}+x_3-2r^2_{2}x_3+r^{4}_{2}x_3)x_{3})\Phi
_{\pi} ( x_{3}
)\Psi^{L}(x_{2})\nonumber \\
&&+r^{2}_{2}(2r_{\pi}\Phi^{t}_{\pi}(x_{3})
-\frac{1}{2}(1-r^{2}_{2})\Phi_{\pi}(x_{3}))\Psi^{t}(x_{2})
 \nonumber \\
&& -r_{\pi} (1-r^2_2)x_3 \Phi^{p}_{\pi}(x_3) \Psi _{L}(x_{2})-
r_{\pi} \left( 2r^{2}_{2}x_2+(1-r^2_2)x_3 \right)
\Phi^{t}_{\pi}(x_3)
\Psi^{L}(x_{2})\Big]\nonumber \\
&& \times E_{1,4}(t^{(2)}_d)
h_d^{(2)}(x_1,x_2,x_3,b_1)\;,\label{psi4}\\
{\cal M}_{6}^{(J/\psi \pi)} &=&16\pi m_B^2 C_{F}\sqrt{2N_{c}}%
\int_{0}^{1}[dx]\int_{0}^{\infty }b_{1}db_{1} \Phi _{B}(x_{1},b_{1})
\nonumber \\
&& \times \Big\{  \Big[
(1-x_2+r^4_{2}x_2+x_3-2r^2_{2}x_3+r^4_{2}x_3-r^4_{2})\Phi _{\pi} (
x_{3} )\Psi^{L}(x_{2})+\nonumber \\
&&r^{2}_{2}(2r_{\pi}\Phi^{t}_{\pi}(x_{3})-\frac{1}{2}(1-r^{2}_{2})\Phi_{\pi}(x_{3}))\Psi^{t}(x_{2})
 \nonumber \\
&& -r_{\pi} (1-r^2_2)x_3 \Phi^{p}_{\pi}(x_3) \Psi^{L}(x_{2})-
r_{\pi} \left( 2r^{2}_{2}(1-x_2)+(1-r^2_2)x_3 \right)
\Phi^{t}_{\pi}(x_3)
\Psi^{L}(x_{2})\Big]\nonumber \\
&&\times E_{6}(t_d^{(1)})h_d^{(1)}(x_1,x_2,x_3,b_1)
\nonumber \\%
&& - \Big[ (1-2r^{2}_{2}+r^{4}_{2})x_{2} \Phi _{\pi} ( x_{3}
)\Psi^{L}(x_{2})
+ \frac{1}{2} (r^{2}_{2}-r^{4}_{2})r^{2}_{2} \Phi_{\pi}(x_{3})\Psi^{t }(x_{2}) \nonumber \\
&&- r_{\pi} (1-r^2_2)x_3 \Phi^{p}_{\pi}(x_3) \Psi^{L}(x_{2})+
r_{\pi} \left( 2r^{2}_{2}x_2+(1-r^2_2)x_3 \right)
\Phi^{t}_{\pi}(x_3) \Psi^{L}(x_{2})
  \Big]\nonumber \\
&& \times E_{6}(t^{(2)}_d) h_d^{(2)}(x_1,x_2,x_3,b_1)
\Big\}\;,\label{psi6}
\end{eqnarray}
with the color factor $C_F=4/3$, the number of colors $N_c=3$, the
symbol $[dx]\equiv dx_1 dx_2 dx_3$ and the mass ratio
$r_{\pi}=m_0^{\pi}/m_B$, $m_0^{\pi}$ being the chiral scale
associated with the $\pi$ meson.

The evolution factor $E_{i}$ and hard function $h_{d}$ in
Eq.(\ref{psi6}) can be found in Ref.~\cite{btocharmn2}. In the
derivation of spectator correction in the pQCD, we need to take the
wave function of relevant mesons, we list the wave functions in
appendix.

For the $B^0$ decay, the CP asymmetry is time dependent,
\begin{eqnarray}
A_{CP}(t)
 &=&
\frac{\Gamma({\bar B}^0(t)\to {J/\psi \pi^0})- \Gamma(B^0(t)\to
{J/\psi \pi^0})} {\Gamma({\bar B}^0(t)\to {J/\psi
\pi^0})+\Gamma(B^0(t)\to {J/\psi\pi^0})}
\;,\nonumber\\
&=& S_{J/\psi \pi^0}\sin(\Delta M t) -C_{J/\psi\pi^0}\cos(\Delta M
t) \;, \label{eq:asymmetry}
\end{eqnarray}

Where the mixing-induced asymmetry $S_{J/\psi\pi^0}$ and direct CP
asymmetry is defined as
\begin{eqnarray}
S_{J/\psi \pi^0}= \frac{2\,{\rm Im}\,\lambda_{J/\psi \pi^0}}
{1+|\lambda_{J/\psi\pi^0}|^2}\;,
\nonumber\\
C_{J/\psi\pi^0}= \frac{1-|\lambda_{J/\psi \pi^0}|^2}
{1+|\lambda_{J/\psi \pi^0}|^2}\;,
\end{eqnarray}
where
\begin{equation}
\lambda_{CP} = \frac{ V_{tb}^*V_{td} \langle J/\psi\pi^0 |H_{eff}|
\overline B^0\rangle} { V_{tb}V_{td}^* \langle J/\psi\pi^0 |H_{eff}|
B^0\rangle}.
\end{equation}

  There are two ways to extract weak phase $\beta$ through $B^0\to
J/\Psi~\pi^0$ decay. The first way is  through  the dependence of
the mixing-induced CP asymmetry on weak phase $\beta$.
   The $S_{J/\psi\pi^0}$ is not sensitive
 of input parameters, as shown in Fig.~\ref{deltaS}.
 That means that the theoretical uncertainties of
 $S_{J/\psi\pi^0}$ is quite small.
 If we measure the mixing-induced asymmetry
$S_{J/\psi\pi^0}$, we can determine weak phase $\beta$  through the
dependence of $S_{J/\psi\pi^0}$ on  $\beta$ as shown in
Fig.~\ref{SandAcp} and Table ~\ref{Tswithbelta},
\begin{center}
\begin{table}[h]
\begin{tabular}{|c|c|c|c|c|c|c|c|c|}
  \hline
 $ \beta$(deg) & 18.0 & 18.3 & 18.6 & 18.9  & 19.2 & 19.5  & 19.8  &
 20.1
  \\ \hline
 $S_{J/\psi\pi^0}$ & -0.58515 & -0.59357 & -0.60192 &-0.61021 & -0.61843 & -0.62658 &
 -0.63467
  & -0.64269 \\ \hline
 $ \beta$ (deg)& 20.4 & 20.7 & 21 & 21.3 & 21.6 & 21.9 & 22.2 & 22.5 \\ \hline
 $S_{J/\psi\pi^0}$& -0.65063 & -0.65851 & -0.66631 & -0.67404 & -0.68170 &-0.68929 & -0.69680 & -0.70424 \\
  \hline
  $ \beta$ (deg)& 22.8 & 23.1 & 23.4 & 23.7 & 24.0 & 24.3 & 24.6 & 24.9 \\ \hline
 $S_{J/\psi\pi^0}$& -0.71160 & -0.71888 & -0.72608 & -0.73321 &-0.74025 &-0.74722 & -0.75410& -0.76090  \\
  \hline
\end{tabular}
\caption{Determination of weak phase $\beta$  through mixing-induced
CP asymmetry $S_{J/\psi\pi^0}$}\label{Tswithbelta}
\end{table}
\end{center}
Another way is to  use   the relation  of the total asymmetry
$A_{CP}$ with the weak phase $\beta$. By integrating $A_{CP}(t)$with
respect to the time variable t, we can get the total asymmetry
$A_{CP}$,
\begin{equation}\label{totalacp}
A_{CP} =  \frac{x}{1+x^2} S_{J/\psi\pi^0}-\frac{1}{1+x^2}
C_{J/\psi\pi^0} ,
\end{equation}
with $x=\Delta m/\Gamma \simeq 0.723 $ for the $B^0$-$\overline B^0$
mixing in the SM \cite{pdg2006}.

Like  the mixing-induced asymmetry, the total asymmetry is also not
sensitive to the input parameters, so we can determine the weak
phase through the relation of  the total CP asymmetry  with weak
phase $\beta$ shown in Fig.~\ref{SandAcp}.

 The numerical calculation needs some parameters and meson
distribution amplitudes as input, we list them in the appendix.

 With the parameters and meson distribution amplitude in the
 appendix, we get the branching ratios of $B\to J/\Psi~ \pi^{0}$
 decays, $\Delta S_{J/\psi\pi^0 }$ and $C_{J/\psi\pi^0 }$,
\begin{eqnarray}
Br(B^0\to J/\psi \pi^0) &=& [1.89^{+ 0.182 }_ {-0.21}(\omega b)^{+
0.0496 }_ {-0.02}(\mu)^{+ 0.193 }_ {-0.171}(F_1)^{+ 0.015 }_
{-0.014}(f_{J/\psi})^{+ 0.04 }_ {-0.059}(\lambda)^{+ 0.04 }_
{-0.068}(A)] \times 10^{-5}
\, ,\nonumber \\
C_{J/\psi\pi^0 } &=& [-9.936 _{- 3.093 }^ {+0.866}(\omega b)_{-
2.368}^{+1.173}(\gamma)_{- 0.289 }^ {+6.914}(\mu)_{- 1.18 }^
{+1.34}(F_1)_{-0.56 }^ {+0.54}(\beta) ] \times 10^{-3}
\, , \nonumber \\
\Delta S_{J/\psi\pi^0 }&=& [2.84 ^{+ 4.07 }_ {-1.00}(\omega b)^{+
0.72 }_ {-0.35}(\gamma)^{+ 2.1 }_ {-0.17}(\mu)^{+ 0.29 }_
{-0.20}(F_1)^{+ 0.03 }_ {-0.05}(\beta)] \times 10^{-3} \, .
\label{BRpipi0}
\end{eqnarray}

The main theoretical errors of the branching ratio are  induced by
the uncertainties below. The first error is from  $\omega
b=0.4\pm0.04 GeV$, the second one is due to  renormalization scale
$\mu$ taken from $mb/2$ to $mb$, the third one is induced by $15\%$
uncertainty of $B \to \pi$ form factor $F_1^{B \to \pi}$, the fourth
one arise from decay constant $f_{J/\psi}=0.405\pm0.05 GeV$, the
fifth error is from CKM matrix parameter $\lambda=0.2272\pm0.001$,
the sixth one is from CKM matrix parameter
$A=0.818^{+0.007}_{-0.017}$.

Compared with the experimental data\cite{pdg2006}
\begin{eqnarray}
Br(B^0 \to J/\psi \pi^0) &=& (2.2 \pm 0.4) \times 10^{-5} \, ,
\label{BRexppipi0}
\end{eqnarray}
our prediction of the branching ratio for $B\to J/\Psi~\pi^{0}$ is
consistent with it.

Unlike the branching ratio, $\Delta S_{J/\psi\pi^0 }$ and
$C_{J/\psi\pi^0 }$   is not sensitive to CKM matrix parameter
$\lambda$ or $A$, because these parameter dependences cancel out.
The independence  of $\Delta S_{J/\psi\pi^0 }$ and $C_{J/\psi\pi^0
}$ on some CKM  parameters is shown in Fig.~\ref{deltaS}(a),(b),and
Fig.~\ref{mixingcp}.(a),(b).

To find new physics and to extract the weak phase $\beta$, we need
 reliable evaluation for the direct CP asymmetry
$C_{J/\psi\pi^0 }$ and $\Delta S_{J/\psi\pi^0}$, so we now consider
the dependence of
 the direct CP
asymmetry $C_{J/\psi\pi^0 }$ and  $\Delta S_{J/\psi\pi^0}$with all
parameters of input.
The main uncertainties of $C_{J/\psi\pi^0 }$ and $\Delta
S_{J/\psi\pi^0 }$ are induced by uncertainties of shape parameter
$\omega b$, CKM matrix phase$\gamma$, renormalization scale $\mu$,
$B \to \pi$ form factor $F_1^{B \to \pi}$ and  the
 weak phase $\beta$. The uncertainties of  $\Delta
S_{J/\psi\pi^0 }$ and  $C_{J/\psi\pi^0 }$ are shown in
Fig.~\ref{deltaS}(c)-(f) and Fig.~\ref{mixingcp}.(c)-(f).

Comparing with the result in Ref.~\cite{deltasciu},
\begin{eqnarray}
  C_{J/\psi\pi^0 }&=& 0.09\pm0.19 \\
S_{J/\psi\pi^0 }&=& -0.47\pm0.30
\end{eqnarray}
 our  results of $\Delta
S_{J/\psi\pi^0
 }$and
$C_{J/\psi\pi^0 }$ has much less theoretical uncertainties. So we
conclude that if the measured deviation $\Delta S_{J/\psi\pi^0 }$ of
the mixing-induced asymmetry is at $1\%$ or the direct asymmetry
$C_{J/\psi\pi^0 }$ is at the level of percentage then we can say
that there should be new physics . We are expecting precise
measurement to the CP asymmetry of $B^0\to J/\psi \pi^0$ in the near
future.

\begin{appendix}
\section{Input Parameters And Wave Functions}
We use the following input parameters in the numerical calculations
\beq
 \Lambda_{\overline{\mathrm{MS}}}^{(f=4)} &=& 250 {\rm MeV}, \quad
 f_\pi = 130 {\rm MeV}, \quad f_B = 190 {\rm MeV}, \non
 m_0^{\pi}&=& 1.4 {\rm GeV},\quad
 M_B = 5.2792 {\rm GeV}, \quad \tau_{B^0}=1.53\times 10^{-12}{\rm
 s},
 \label{para}
\eeq
   For the CKM matrix elements,  we adopt the wolfenstein
parametrization for the CKM matrix up to $\mathcal{O}$$(\lambda^
3)$\cite{pdg2006},
\begin{equation}
V_{CKM}= \left(           \begin{array}{ccc}
          1-\frac{\lambda^2}{2} & \lambda & A \lambda^3 (\rho-i \eta)\\
          -\lambda & 1-\frac{\lambda^2}{2} & A \lambda^2 \\
          A \lambda^3 (1-\rho-i \eta )&-A \lambda^2 & 1
          \end{array} \right) , \label{vckm}
\end{equation}
with the parameters $\lambda=0.2272, A=0.818, \rho=0.221$ and
$\eta=0.340$.

For the $B$ meson distribution amplitude, we adopt the
model\cite{kls01}

\beq \phi_B(x,b) &=& N_B x^2(1-x)^2 \mathrm{exp} \left
 [ -\frac{M_B^2\ x^2}{2 \omega_{b}^2} -\frac{1}{2} (\omega_{b} b)^2\right],
 \label{phib}
\eeq

where $\omega_{b}$ is a free parameter and we take
$\omega_{b}=0.4\pm 0.05$ GeV in numerical calculations, and
$N_B=91.745$ is the normalization factor for $\omega_{b}=0.4$.

The $J/\psi$ meson asymptotic distribution amplitudes are given by
\cite{BC04}
\begin{eqnarray}
\Psi^L(x)&=&\Psi^T(x)=9.58\frac{f_{J/\psi}}{2\sqrt{2N_c}}x(1-x)
\left[\frac{x(1-x)}{1-2.8x(1-x)}\right]^{0.7}\;,\nonumber\\
\Psi^t(x)&=&10.94\frac{f_{J/\psi}}{2\sqrt{2N_c}}(1-2x)^2
\left[\frac{x(1-x)}{1-2.8x(1-x)}\right]^{0.7}\;,\nonumber\\
\Psi^V(x)&=&1.67\frac{f_{J/\psi}}{2\sqrt{2N_c}}\left[1+(2x-1)^2\right]
\left[\frac{x(1-x)}{1-2.8x(1-x)}\right]^{0.7}\;,\label{jda}
\end{eqnarray}

For the light meson wave function, we neglect the $b$ dependant
part, which is not important in numerical analysis. We choose the
wave function of $\pi$ meson \cite{ball3}:
\begin{eqnarray}
 \Phi_\pi(x) &=&  \frac{3}{\sqrt{6} }
  f_\pi  x (1-x)  \left[1+0.44C_2^{3/2} (2x-1) +0.25 C_4^{3/2}
  (2x-1)\right],\label{piw1}\\
 \Phi_{\pi}^P(x) &=&   \frac{f_\pi}{2\sqrt{6} }
   \left[ 1+0.43 C_2^{1/2} (2x-1) +0.09 C_4^{1/2} (2x-1) \right]  ,\\
 \Phi_{\pi}^t(x) &=&  \frac{f_\pi}{2\sqrt{6} } (1-2x)
   \left[ 1+0.55  (10x^2-10x+1)  \right]  .    \label{piw}
 \end{eqnarray}
 The Gegenbauer polynomials are defined by
 \begin{equation}
 \begin{array}{ll}
 C_2^{1/2} (t) = \frac{1}{2} (3t^2-1), & C_4^{1/2} (t) = \frac{1}{8}
 (35t^4-30t^2+3),\\
 C_2^{3/2} (t) = \frac{3}{2} (5t^2-1), & C_4^{3/2} (t) = \frac{15}{8}
 (21t^4-14t^2+1).
 \end{array}
 \end{equation}

\end{appendix}
\hspace{10pt}

\begin{figure}[htb]
\begin{center}
\epsfig{file=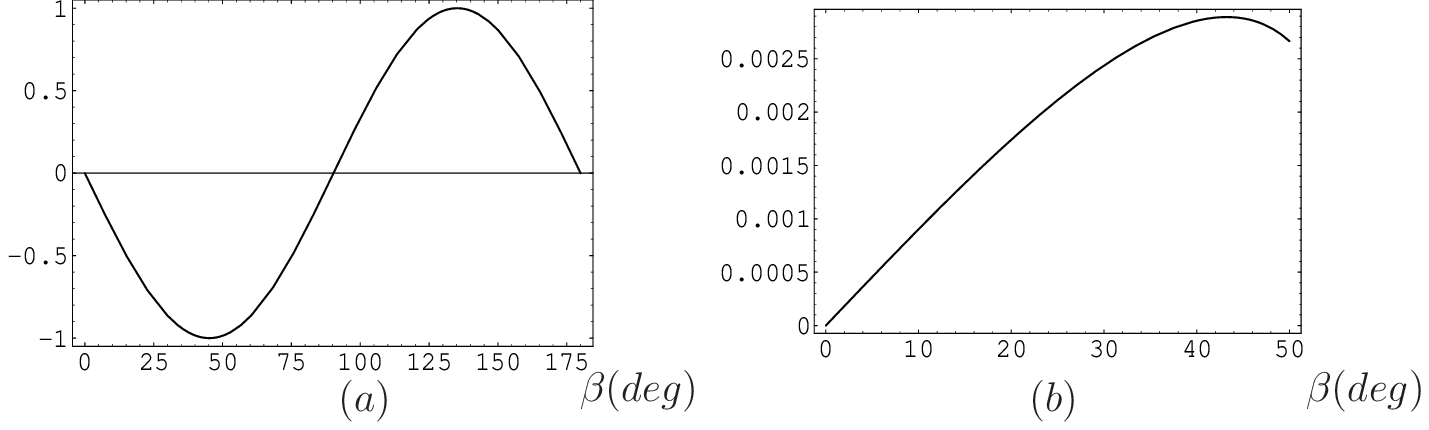,height=4cm,width=14cm}~~~~~~~~
 \caption{The dependence of the mixing-induced asymmetry $
S_{J/\psi\pi^0 }$ for  $B^0\to J/\Psi~\pi^0$ on the weak phase
$\beta$ in diagram  $(a)$. The dependence of the deviation $\Delta
S_{J/\psi\pi^0 }$ of the mixing-induced asymmetry from
$\sin(-2\beta)$ on the weak phase $\beta$ in diagram  $(b)$
}\label{swithbelta} \vskip0.5cm
\psfig{file=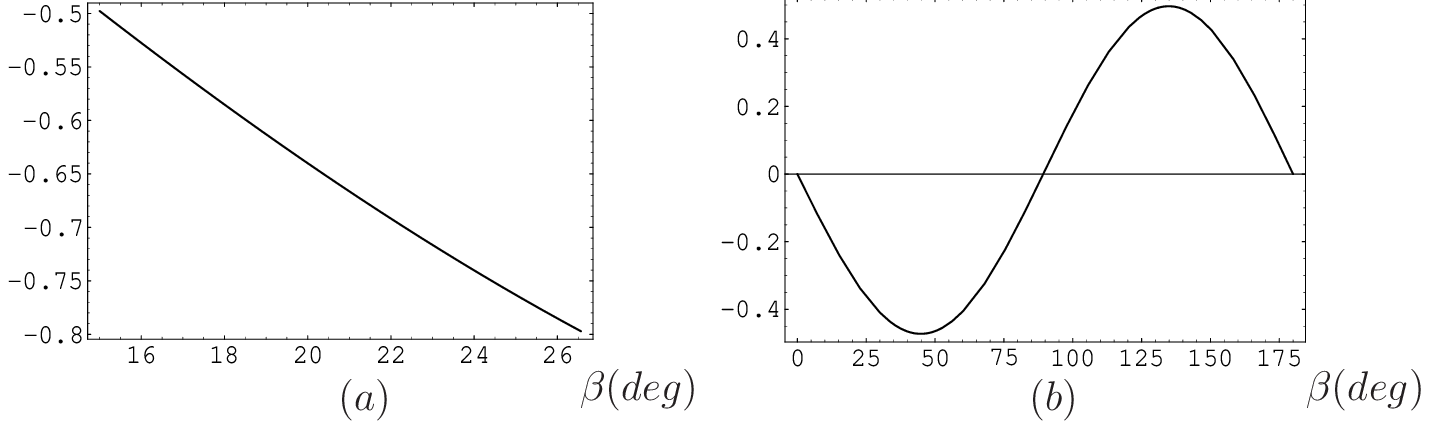,height=4cm,width=14cm}~~~~~~~~
 \caption{The dependence of the the mixing-induced asymmetry $
S_{J/\psi\pi^0 }$ for  $B^0\to J/\Psi~\pi^0$ on the weak phase
$\beta$ in diagram $(a)$ can be used to extract the weak phase
$\beta$ . The dependence  of total CP asymmetry $A_{CP}$ on the weak
phase $\beta$ in diagram$(b)$ can be used to extract the weak phase
$\beta$ also.}\label{SandAcp}
\end{center}
\end{figure}

\vskip1cm
\begin{figure}
\begin{center}
\psfig{file=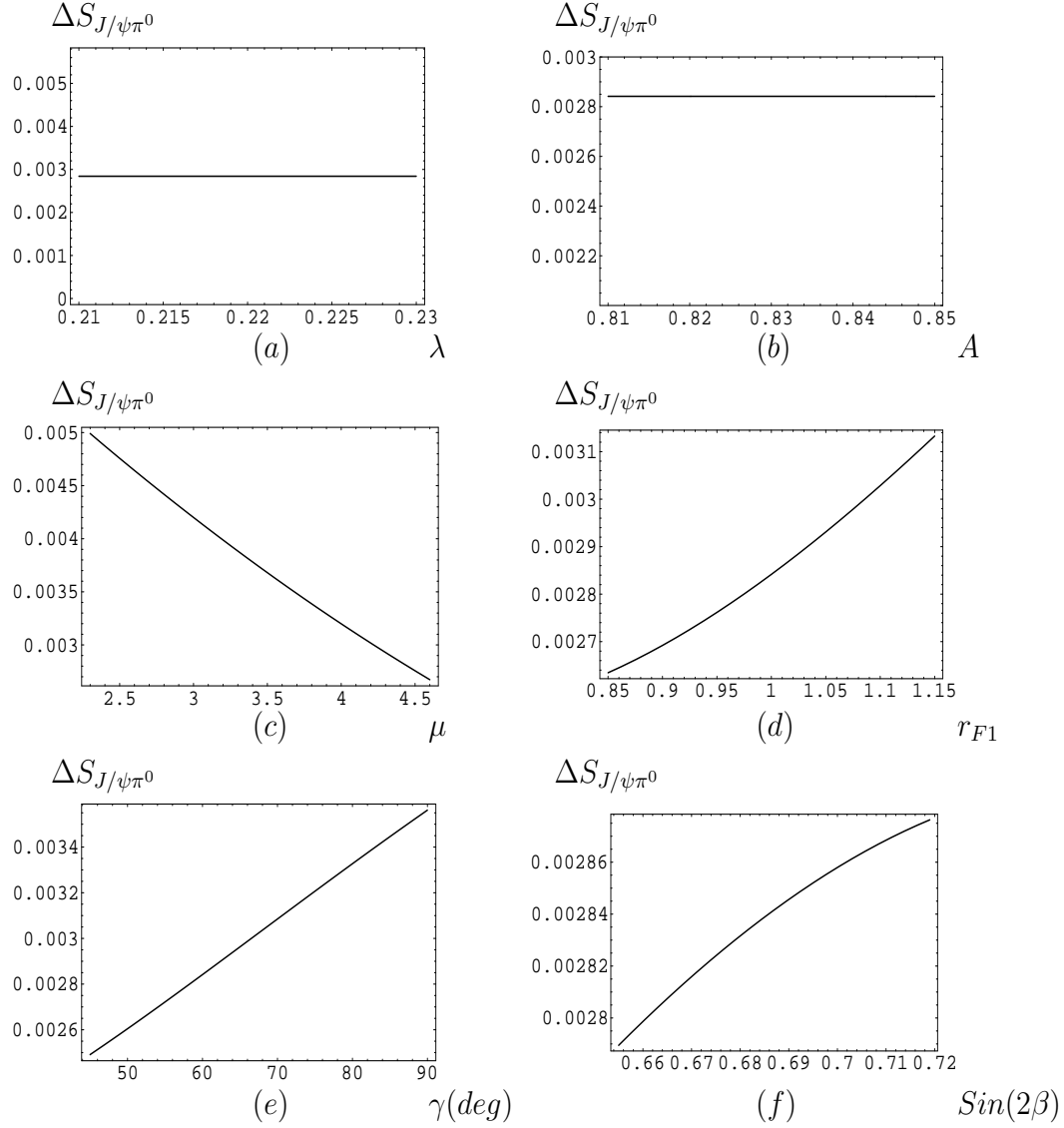,height=15cm,width=14cm}~~~~~~~~
 \caption{The uncertainties of  $\Delta S_{J/\psi\pi^0 }$ of the mixing-induced asymmetry
from $\sin(-2\beta)$  are induced by  that  of renormalization scale
$\mu$ in $(c)$ , that of $B \to \pi$ form factor in $(d)$, that of
the weak phase $\gamma$  in $(e)$ and that of  $\sin(2\beta)$  in
$(f)$ .} \label{deltaS}
\end{center}
\end{figure}




\begin{figure}[h]
\begin{center}
\psfig{file=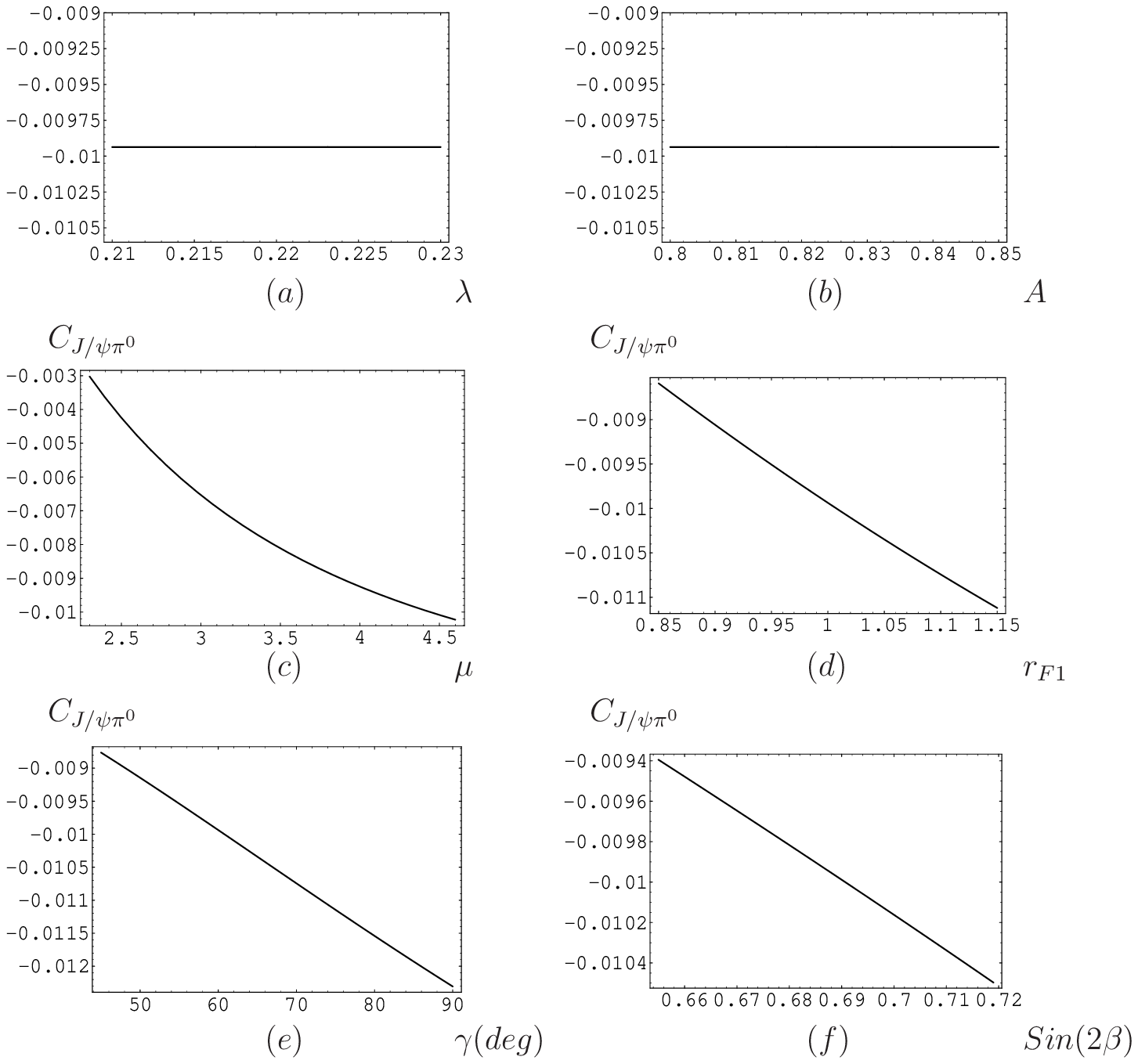,height=15cm,width=14cm}~~~~~~~~
 \caption{The uncertainties of  the direct CP asymmetry $C_{J/\psi\pi^0 }$
 are induced by  that  of renormalization scale
$\mu$ in $(c)$ , that of $B \to \pi$ form factor in $(d)$, that of
the weak phase $\gamma$  in $(e)$ and that of  $\sin(2\beta)$  in
$(f)$ .}\label{mixingcp}
\end{center}
\end{figure}

\begin{thebibliography}{99}

\bibitem{npa1}Gautam Bhattacharyya, Gustavo C. Branco, Wei-Shu Hou,
Phys.Rev. D\textbf{54}(1996)2114; P. Bamert, Int.J.Mod.Phys.
A\textbf{12}(1997)723; G.F.Giudice, M.L.Mangano, et al,
hep-ph/9602207.
 \bibitem{npa2}P. Bamert, C.P. Burgess, J.M. Cline, D. London, E.
 Nardi,Phys.Rev. D\textbf{54} (1996) 4275;
 J.L. Hewett, T. Takeuchi, S. Thomas
,hep-ph/9603391;
 J.-M. Fr¨¨re, V.A. Novikov, M.I. Vysotsky,
Phys.Lett. B\textbf{386} (1996) 437.
\bibitem{npa3} F. Larios, C.-P. Yuan, Phys.Rev. D\textbf{55 }(1997)
7218;
 M. Gronau, D. London,Phys.Rev. D\textbf{55} (1997)
2845;
 Joao P. Silva, L. Wolfenstein, Phys.Rev. D\textbf{55} (1997)
5331.
\bibitem{npa4}Dongsheng Du, Hongying Jin, Yadong Yang, Phys.Lett. B\textbf{414} (1997)
130; S.I.Bityukov, N.V.Krasnikov, Mod.Phys.Lett. A\textbf{12} (1997)
2011; Robert Fleischer, Thomas Mannel, hep-ph/9706261; A.I. Sanda,
Zhi-zhong Xing, Phys.Rev. D\textbf{56} (1997) 6866.
\bibitem{npa5} A.L. Kagan, M. Neubert, Phys.Rev. D\textbf{58} (1998)
094012; Xiao-Gang He, Wei-Shu Hou, Phys.Lett. B\textbf{445} (1999)
344; Yue-Liang Wu, Chin.Phys.Lett. \textbf{16 }(1999) 339.
\bibitem{npa6}Katri Huitu, Cai-Dian Lu, Paul Singer, Da-Xin Zhang
, Phys.Rev.Lett. \textbf{81} (1998) 4313; Harry J. Lipkin, Zhi-zhong
Xing, Phys.Lett. B\textbf{450} (1999) 405; Maxime Imbeault, David
London, Chandradew Sharma, Nita Sinha, Rahul Sinha, hep-ph/0608169.
\bibitem{npa7}S. Fajfer, S. Prelovsek, P. Singer, D. Wyler, Phys.Lett. B\textbf{487} (2000)
81; T. M. Aliev, A. Ozpineci, M. Savci, Phys.Rev. D\textbf{65
}(2002) 115002; Cheng-Wei Chiang, Jonathan L. Rosner, Phys.Rev.
D\textbf{68} (2003) 014007; M. Ciuchini, E. Franco, F. Parodi, V.
Lubicz, L. Silvestrini, A. Stocchi, hep-ph/0307195.
\bibitem{npa8}Andrzej J. Buras, hep-ph/0402191; A.K.Giri, R.Mohanta, Phys.Lett. B\textbf{594} (2004)
196; A.K.Giri, R.Mohanta, JHEP \textbf{0411} (2004) 084; Seungwon
Baek, JHEP \textbf{0607} (2006) 025 .
\bibitem{npa9}Rahul Sinha, Basudha Misra, Wei-Shu Hou, Phys.Rev.Lett. \textbf{97} (2006) 131802
; M. Bona, M. Ciuchini, E. Franco, et al, Phys.Rev.Lett. \textbf{97}
(2006) 151803; Hsiang-nan Li, Satoshi Mishima, hep-ph/0610120 ; C.
S. Kim, Sechul Oh, Yeo Woong Yoon, arXiv:0707.2967.
\bibitem{bmixing1} M. Gronau, D. London, Phys.Rev. D\textbf{55 }(1997) 2845.
\bibitem{bmixing2} D. London, hep-ph/9907311.
\bibitem{bmixing3}Patricia Ball, Robert Fleischer, Eur.Phys.J. C\textbf{48} (2006)
413; Patricia Ball, hep-ph/0703214.
\bibitem{bmixingatt1}Alakabha Datta, Phys.Rev. \textbf{D74 }(2006) 014022.
\bibitem{bmixingatt2}Alakabha Datta, Phys.Rev. \textbf{D74} (2006) 014022.
\bibitem{bmixingatt3}Joao P. Silva, Lincoln Wolfenstein, Phys.Rev. \textbf{D62} (2000)
014018.
\bibitem{bmixingatt4} Zhi-zhong Xing, Eur.Phys.J.\textbf{ C4 }(1998) 283-287.
\bibitem{bmixingatt5} George W.S. Hou, hep-ph/0611154.
\bibitem{deltasciu}M. Ciuchini, M. Pierini, L. Silvestrini, Phys.Rev.Lett. \textbf{95} (2005)
221804.
\bibitem{btocharmn1}Chuan-Hung Chen, Hsiang-Nan Li, Phys.Rev. D\textbf{71 }(2005)
114008.
\bibitem{btocharmn2} Jing-Wu Li, Dong-Sheng Du, Phys. Rev. D \textbf{78}, (2008) 074030.
\bibitem{pdg2006}W.-M. Yao et al., Journal of Physics, G 33, 1
(2006).
\bibitem{formbtopi}Patricia Ball, Roman Zwicky, Phys.Rev. D\textbf{71} (2005)
014015.
\bibitem{qcdf} M.~Beneke, G.~Buchalla, M.~Neubert and C.~T.~Sachrajda,
  Phys.\ Rev.\ Lett.\  {\bf 83}, 1914 (1999)
  ;
  Nucl.\ Phys.\ B {\bf 591}, 313 (2000)
  ;
  M.~Beneke, G.~Buchalla, M.~Neubert and C.~T.~Sachrajda,
  Nucl.\ Phys.\ B {\bf 606}, 245 (2001)
  ;
  M.~Beneke and M.~Neubert,
  Nucl.\ Phys.\ B {\bf 675}, 333 (2003)
  .

\bibitem{kls01}
Y.-Y.~Keum, H.-n.~Li and A.I.~Sanda, Phys.Lett. B, {\bf 504},
6(2001) ;
 Phys.Rev. D,{\bf 63}, 054008 (2001) .
 \bibitem{BC04} A.E. Bondar and V.L. Chernyak, hep-ph/0412335.
\bibitem{ball3}
V.M.~Braun and I.E.~Filyanov, Z.Phys.C {\bf 48}, 239 (1990);
P.~Ball, J.High Energy Phys. \textbf{01},010 (1999) .
\end{thebibliography}
\end{document}